\documentclass[aps,prl,twocolumn,showpacs,floatfix,superscriptaddress]{revtex4}
\usepackage{graphicx}

\begin{document}
\flushbottom

\title{Gap opening with ordering in  PrFe$_4$P$_{12}$ studied
by local tunneling spectroscopy}
\author{H. Suderow}
\affiliation{Laboratorio de Bajas Temperaturas, Departamento de F\'isica de la Materia Condensada \\ Instituto de Ciencia de Materiales Nicol\'as Cabrera, Facultad de Ciencias \\ Universidad Aut\'onoma de Madrid, 28049 Madrid, Spain}
\author{K. Behnia}
\affiliation{Laboratoire Photons et Mati\`{e}re (UPR5-CNRS), ESPCI,
10 Rue de Vauquelin, 75231 Paris, France}
\author{I. Guillamon, V. Crespo, S. Vieira}
\affiliation{Laboratorio de Bajas Temperaturas, Departamento de F\'isica de la Materia Condensada \\ Instituto de Ciencia de Materiales Nicol\'as Cabrera, Facultad de Ciencias \\ Universidad Aut\'onoma de Madrid, 28049 Madrid, Spain}
\author{D. Kikuchi}
\affiliation{Department of Physics, Tokyo Metropolitan University,
Tokyo 192-0397, Japan}
\author{Y. Aoki}
\affiliation{Department of Physics, Tokyo Metropolitan University,
Tokyo 192-0397, Japan}
\author{H. Sugawara}
\affiliation{Faculty of Integrated Arts and Science, University of
Tokushima, Tokushima 770-8502, Japan}
\author{ H. Sato}
\affiliation{Department of Physics, Tokyo Metropolitan University,
Tokyo 192-0397, Japan}

\begin{abstract}
We present measurements of the local tunneling density of states in
the low temperature ordered state of PrFe$_4$P$_{12}$. The
temperature dependencies of the Fermi level density of states, and
of the integrated density of states at low bias voltages, show
anomalies at $T\simeq6.5 K$, the onset of multipolar ordering as
detected by specific heat and other macroscopic measurements. In the
ordered phase, we find a local density of states with a V-shape form,
indicating partial gap opening over the Fermi surface. The size of the gap according to the tunneling
spectra is about 2 meV.
\end{abstract}

\pacs{71.27. a+,71.30 +h, 73.20.-r} \date{\today} \maketitle

Recently, a new family of heavy fermion f-electron intermetallic
compounds with the filled skutterudite structure RT$_4$X$_{12}$ (R
is a rare earth, T a transition metal and X a pnictogen) has been
found to show many attractive phenomena, most of them appearing at
low temperatures. Among others, metal-insulator transition, heavy
fermion behavior, and superconductivity have been found in (R=Pr)
systems\cite{Aoki05,Bauer02,Sekine97,Measson04,Suderow04,Sugawara02}.

In particular, in PrFe$_4$P$_{12}$, a peculiar phase transition
occurs at T$_A$ = 6.5 K, with distinct anomalies in both
resistivity\cite{Sato00} and specific heat\cite{Aoki02}. The
intriguing order below T$_A$ is widely believed to be associated
with orbital degrees of freedom\cite{Aoki05}. A recent NMR
study\cite{Kikuchi07} has put severe restrictions on the symmetry of
the order parameter. According to it, the ordering involves
non-magnetic multipoles, which do not break the point symmetry of
the crystal at the Pr sites. The scalar order with
$\Gamma_{1g}$\cite{Kiss06} is currently the most promising candidate
to represent the order parameter. On the other hand, according to
transport studies\cite{Sato00,Sugawara02,Pourret06}, the itinerant
electrons are profoundly affected by this phase transition. The
multifold increase in the Hall, Seebeck and Nernst coefficients all
suggest a drastic decrease in the number of charge
carriers\cite{Pourret06}. This is compatible with the results of a
de Haas van Alphen study\cite{Sugawara02}, which found a small Fermi
surface (0.0015 of the Brillouin zone) with a moderate effective
mass ($\sim$10m$_{e}$) in the ordered state. In higher magnetic
fields (strong enough to destroy the ordering), the same study
resolved several larger Fermi surfaces with heavier masses up to
81m$_{e}$ \cite{Sugawara02}. The consequences of the ordering on
transport properties recall the case of
URu$_2$Si$_2$\cite{Behnia05}. In both cases, the ordering leads to
an enhanced lattice thermal conductivity, a large Hall coefficient
and anomalously large Nernst and Seebeck coefficients due to an
incomplete metal-insulator transition, producing a heavy-Fermion
semi-metal. In the case of URu$_2$Si$_2$, a partial gap opening has
been observed using different techniques\cite{Flouquet05,Naidyuk98}.

On the other hand, scanning tunneling microscopy and spectroscopy (STM/S)
techniques have been widely applied to systems where some kind of energy gap
appears in the quasiparticle spectrum. The superconducting gap is certainly
one of the most studied kinds of energy gaps viewed using this technique, although
different kinds of charge ordered, and semimetallic or semiconducting compounds,
have also received attention\cite{Fischer07,Hofer03}. Any features in the density
of states at energies of the order of the meV are relatively easy to resolve using
STM/S, because the tip density of states remains featureless and temperature independent below several tens of meV.
Therefore, when a tip is scanned over a surface, the bias dependence
of the local tunneling conductance, below some tens of mV, is simply the temperature smeared local density of states of the sample.
Examples of STM/S experiments showing a temperature dependent gapped density of states
include e.g. the intriguing pseudogap of the High T$_c$ superconductors,
the charge density wave gap in the di-chalchogenides, or the peculiar low
energy electronic features appearing in graphite \cite{Fischer07,Coleman88,Matsui05}.
Here we present local tunneling conductance measurements at low temperatures made
with STM/S that evidence the appearance of a gap in the quasiparticle spectrum of
the heavy fermion material PrFe$_4$P$_{12}$.

\begin{figure}[ht]
\includegraphics[width=8cm,clip]{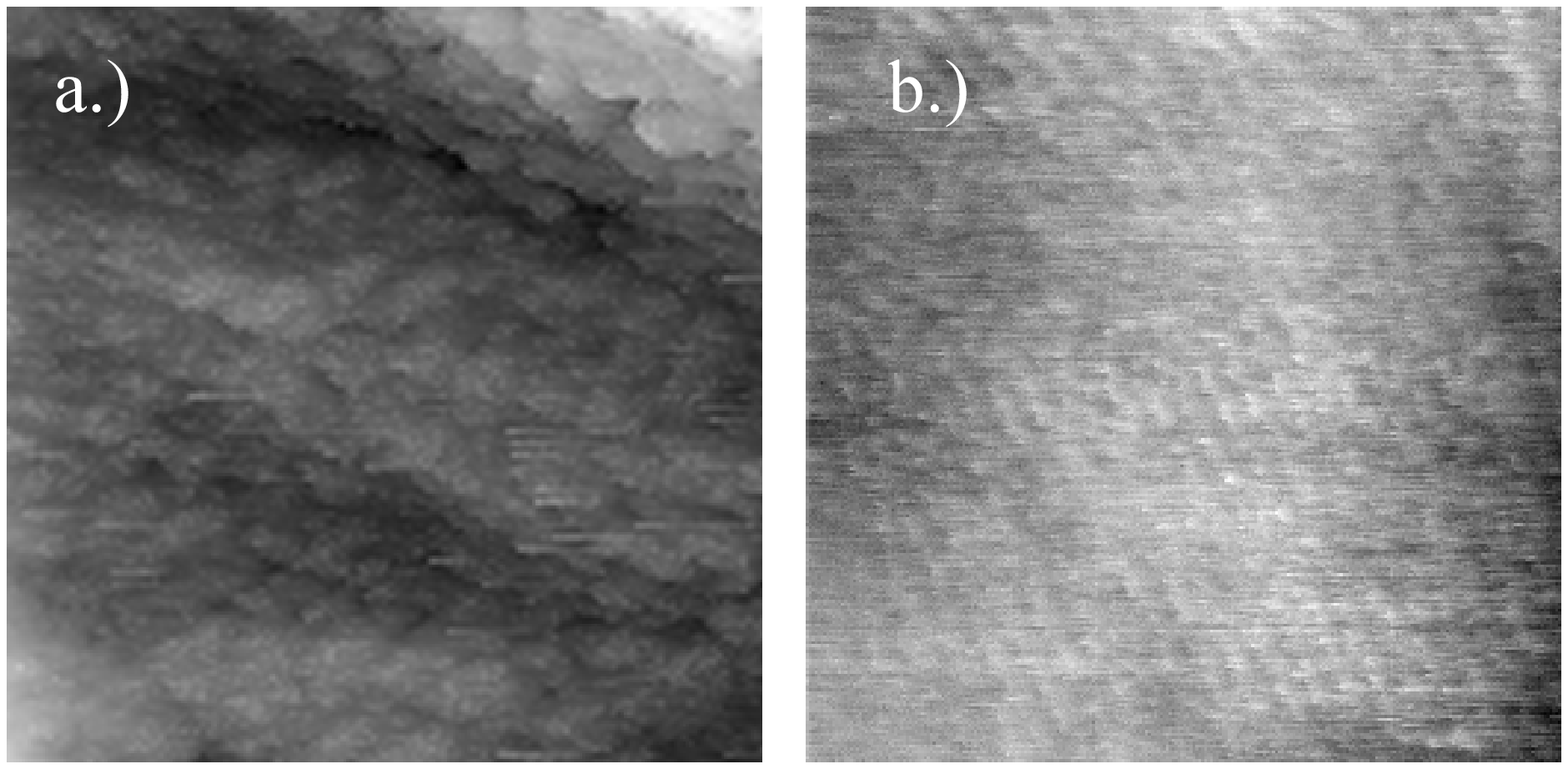}
\vskip -7cm \caption{In a. we show a 90 nm x 90 nm region with small irregular terraces, and in b. an image of 7 nm x 7 nm, where lines at distances of around 0.2 nm, i.e. of the same order as interatomic distances of the unit cell, are observed. Grey scale of the images accounts for height differences of, respectively, 13 nm in a. and 0.5 nm in b. The corrugation of the terraces and steps observed in a. is of about 2 nm, and the corrugation of the linear structures observed in b. is of about 0.1 nm. A plane has been subtracted from both images. They have been acquired at 1.5 K, on a sample of PrFe$_4$P$_{12}$ with the surface perpendicular to the (100) direction. The tunneling density of states is reproducible over the whole surface, always showing the behavior presented in the Figs. \protect\ref{Fig2} and \protect\ref{Fig3}.} \label{Fig1}
\end{figure}

We use a home built STM/S system in a $^4$He refrigerator that goes down to 1.5 K, and has been used for several previous
work in superconducting materials\cite{Suderow01,Rubio01,Rodrigo04b}. The STM/S has a scanning range of about
2 x 2 $\mu$m at low temperatures, and a sample holder that allows to move in-situ the tip over the whole area of
the sample. When needed, the tip is prepared and cleaned in-situ following the methods discussed
in Refs.\cite{Rodrigo04b,Guillamon07}. We took a small single crystalline sample broken at ambient conditions
and mounted into the STM sample holder, and used tips of Au. The tip tunnels into the (100) direction of the single
crystalline sample. The sample showed large surface areas where clean tunneling conditions could be achieved,
with tunneling conductance curves that were reproducible over the whole surface. Moreover, the current-distance characteristics signalled work functions of several eV, and the topography and spectroscopy was independent of the tunneling resistance.

\begin{figure}[ht]
\includegraphics[width=8cm,clip]{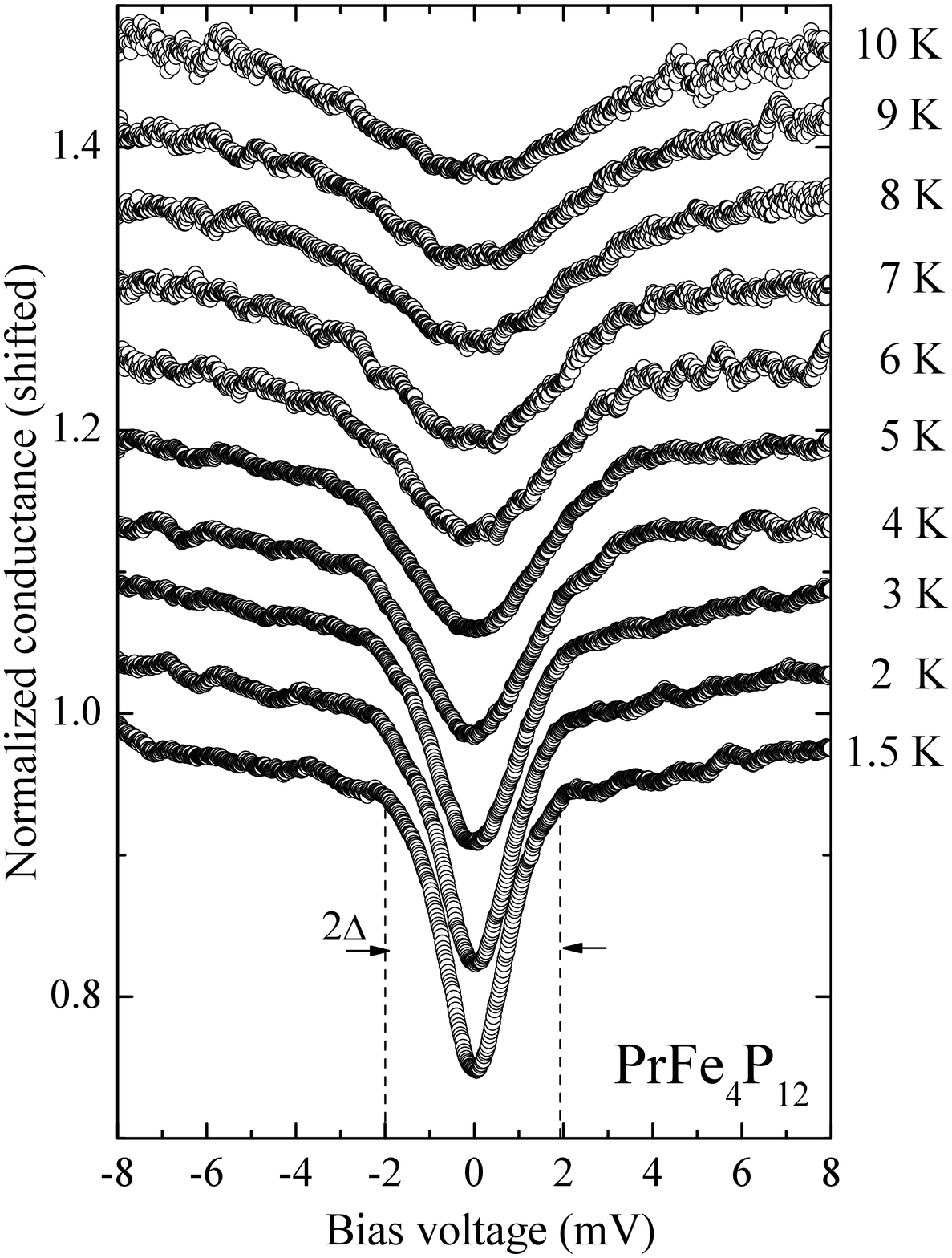}
\vskip -0.5cm \caption{The temperature dependence of the local tunneling density of states of PrFe$_4$P$_{12}$ between 1.5K and 10 K. The conductance is of 1$\mu$S, and the data have been normalized to 10 mV and displaced by 0.05 in the y axis for clarity (temperature for each curve is given on the right). We find a steep decrease of the density of states below about 2 mV that develops below 7 K. The size of the low temperature gap is estimated from the onset of the decrease in the conductance at low temperatures, as schematically described by the dashed lines.} \label{Fig2}
\end{figure}

Typical images (Figs.\ref{Fig1}) consist of irregular terraces or
small nanoscopic size granular structures, very similar to the ones
observed on the surface of several superconducting materials
\cite{Suderow01,Rubio01,Suderow04}. In some surface regions, we
could observe regular terraces with characteristic distances of the
order of the unit cell (Fig.\ref{Fig1}b). Over the whole surface, we
reproducibly observe a gapped local density of states. The curves
shown in Fig.\ref{Fig2} are obtained in large images of the surface,
above and below the ordering temperatures. Above T$_A$, the
tunneling conductance shows no significant temperature dependence,
besides temperature smearing. The curves are usually slightly
parabolic, with a density of states, which has a small bias
dependence around the Fermi level. Typically, in this voltage range,
metals show a flat density of states. Here, the small decrease of
the density of states may signal a band structure with a peculiar
form at energies very close to the Fermi level, that may preclude
the opening of the gap in the ordered phase. When ordering sets in,
a new energy scale appears in the tunneling conductance curves,
which show a steep decrease for voltages below about 2 mV. At the
lowest temperatures (1.5 K), the decrease is sharpest. The local
density of states shows then closely a linear behavior at very low
energies, with a characteristic V shape and a density of states at
the Fermi level that remains relatively high.

\begin{figure}[ht]
\includegraphics[width=8cm,clip]{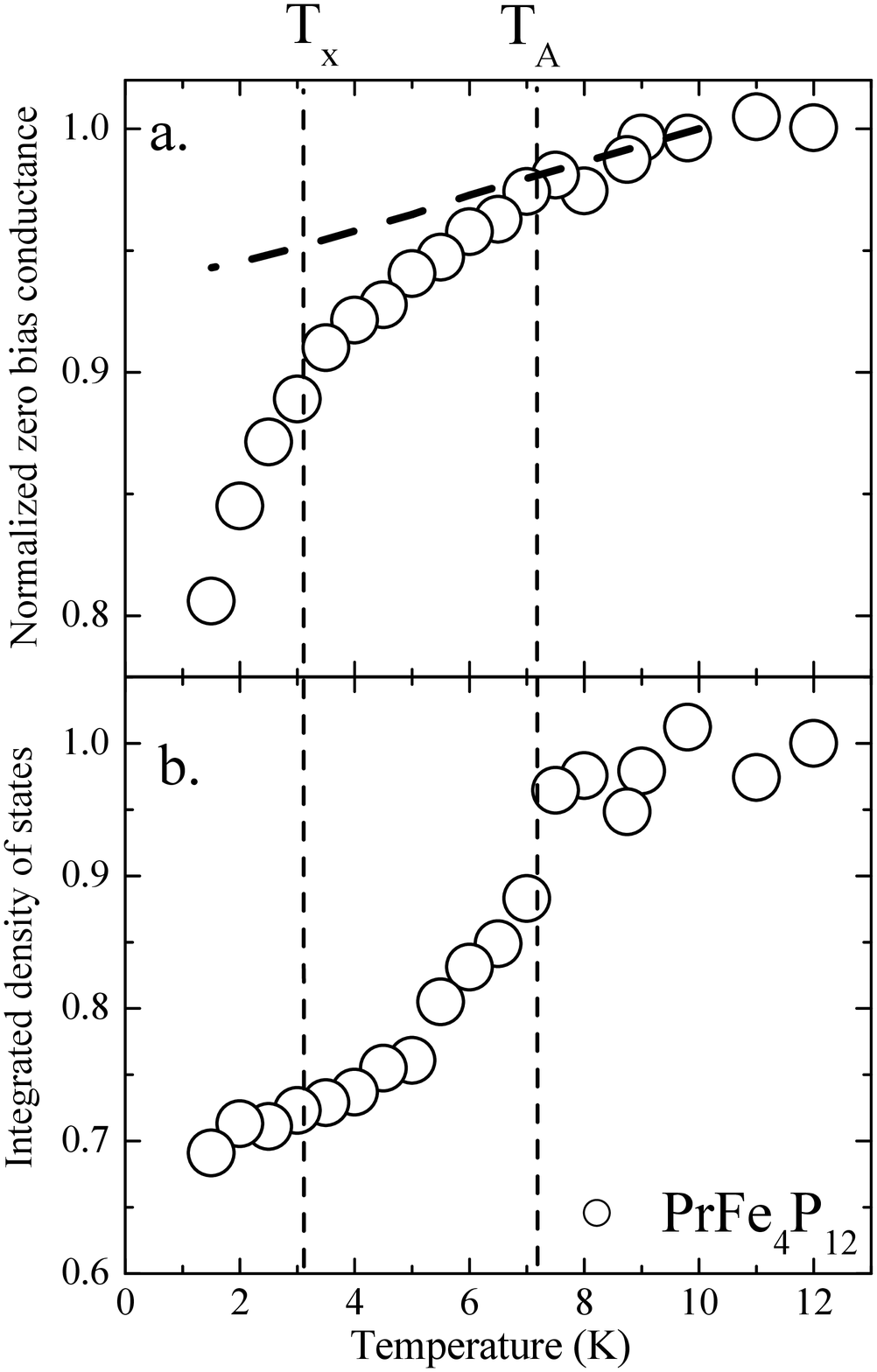}
\vskip -0cm \caption{In a. we show the zero bias conductance as a function of temperature. Vertical lines represent the temperatures of onset of ordering T$_A$ and T$_x$, obtained in macroscopic measurements (see text). Dashed line represents the extrapolation of the zero bias conductance from the data above T$_A$ down to lower temperatures, calculated by gradually eliminating thermal smearing. The Fermi level density of states shows, at each one of both ordering temperatures, a clear decrease. In b. we show the integral over the tunneling conductance normalized to 1 above T$_A$.}
\label{Fig3}
\end{figure}

Note that the decrease is not a simple consequence of the reduced
temperature smearing of the data above T$_A$. To see this, we have
plotted as the dashed line in Fig.\ref{Fig3}a the zero bias
conductance calculated by reducing temperature smearing from the
tunneling conductance above T$_A$, and normalized the result to one
at 10 K. The expected variation is smooth. However, the
experimental data (circles in Fig.\ref{Fig3}a) show a steep decrease
which sets in at T$_A$ and becomes more pronounced below T$_x$. The
latter temperature scale (T$_x\simeq2.5 K$) is associated with features in the transport data (most visible in the Hall
mobility and the Nernst coefficient\cite{Pourret06}). These two
characteristic temperatures are recovered as kinks in the
temperature dependence of the zero bias conductance. When
extrapolating these data to zero temperature, we find that the local
Fermi level density of states is about 30\% smaller than in the high
temperature phase. Furthermore, the integral over the local density
of states measured at low energies (Fig.\ref{Fig3}b) also changes
strongly below T$_A$. Indeed, the integral of tunneling conductance
curves obtained from a temperature independent density of states
also remains temperature independent. Here we observe instead a
strong drop below T$_A$, evidencing the loss of charge carriers
associated with the ordered state. Actually, the integral rapidly
falls down to about 70\% its value in the high temperature phase,
and remains roughly constant, even on crossing the transition T$_x$.
So below T$_x$, the decrease of the local Fermi level density of
states (Fig.\ref{Fig3}a), is compensated by a slight increase at the shoulders of the conductance curves around 2mV.

The gap edge can be obtained from the voltage position of the onset of the decrease of the tunneling conductance curves down to the V shape observed at low temperatures (dashed lines in Fig.\ref{Fig2}). We find $\Delta$ = 2 meV, which gives about 2$\Delta$/k$_B$T$_A$$\approx$7, a value that is significantly larger than the mean field BCS result (2$\Delta$/k$_B$T$_A$ = 3.7) previously used in the discussion of charge or spin density wave phenomena\cite{Coleman88,Dai93,Kiss07}.

On the other hand, the origin of the V-shape density of states and
the significance of the high density of states obtained at the Fermi
level and at low temperatures have to be discussed more carefully.
As it is well known, in a local tunneling spectroscopy experiment,
the obtained local density of states is formed by contributions
coming from very different parts of the Fermi
surface\cite{Fischer07}. There is no selectivity in the direction
perpendicular to the surface of the sample, as in a planar junction,
and, in principle, many in-plane and out of plane momenta are
probed\cite{Fischer07,Hofer03}. However, precise information of the
actual electrons taking part in the tunneling process is very
difficult to obtain without atomic resolution and comparison with
theoretical calculations\cite{Guillamon07}. For instance, note that
in the well studied superconducting material MgB$_2$, where until
now no STM/S experiment has been made with atomic resolution, nearly
all STM/S data show the opening of the superconducting gap on only
part of the Fermi surface, namely the so called $\pi$ band
gap\cite{Rubio01}. Data from the rest of the Fermi surface, the so
called $\sigma$ bands, where the electron phonon coupling is larger,
are more difficult to obtain using STM/S. Actually, STM/S
experiments in this compound have never unambiguously shown
tunneling features close to what one would expect taking into
account the opening of the gap over the whole Fermi
surface\cite{Suderow02b,Eskildsen03,VolumePhysC}. In general, it is
reasonable to assume that lighter electrons couple better with the
states of the tip, so that local tunneling may preferentially
reflect the properties of the Fermi surface with less mass
renormalization, as possibly occurs in MgB$_2$.

In the case of  PrFe$_4$P$_{12}$,  the magnitude of decrease in the
conductance integral indicates that about one third of the electrons
participating in the tunneling process come from the gapped part of
the Fermi surface. The rest, i.e. the vast majority, must come from
part of the Fermi surface where there is no gap at all, or it is
vanishingly small. The de Haas van Alphen data\cite{Sugawara02} have
detected a small and almost spherical Fermi surface in the center of
the Brillouin zone. The size of this Fermi surface is in very good
agreement with the carrier density extracted from the
low-temperature positive Hall coefficient. Therefore, it is
reasonable to assume that this is a hole-like band with a large
electronic mobility dominating the Hall response of the system.
However, other Fermi surfaces, electron-like, with a larger mass and
a lower mobility may have remained undetected by dHvA. They would
account for the relatively large magnitude of the linear specific
heat\cite{Sugawara02} and thermopower\cite{Pourret06}. According to
band structure calculations\cite{Harima02}, in addition to a small
hole-like spherical surface at the $\Gamma$-point, there is a large quasi-cubic
Fermi surface subject to strong a nesting instability. Nesting would
lead to a quasi-complete destruction of this larger Fermi surface,
leading to small surviving portions (with unknown shape and energy
dispersion) coexisting with the small spherical Fermi surface at the
center of the Brillouin zone.

Thus, it is tempting to assume that local tunneling spectroscopy
experiments preferentially probe the small Fermi surface barely
affected by the ordering. The large Fermi level conductance found at
low temperatures may result from this Fermi surface. On the other
hand, the opening of the gap and the V-shape structure of the curves,
may be a signature of the other (that is the larger, nested and
undetected) Fermi surface. This would explain the non-observation of
a fully developed gap in our experiments, as may be naively expected taking into
account the reduction of charge carriers in the ordered phase
observed in other experiments\cite{Sugawara02,Pourret06}, and would be also
compatible with the nesting scenario.

We note also that the temperature dependence of the local tunneling
conductance shown in Fig.\ref{Fig3} suggests that the behavior
observed here is generated by the bulk and is not merely a surface
property. Nevertheless, as in all STM/S experiments, surface related
effects could in principle influence some of the features of the
spectra, as their precise form or voltage dependence.

In summary, we have detected the consequences of the ordering on
local tunneling spectroscopy. The temperature dependence of the
tunneling spectra coincides with bulk results. Thus, the measured
tunneling density of states reproduces the behavior found in part of
the Fermi surface. In particular, we find a gap is of 2 meV, which
gives 2$\Delta/k_BT_c\approx$ 7, and that there are still parts of
the Fermi surface that remain ungapped. This shows that the peculiar
ordering in this compound leads to a largely frustrated metal
insulator transition, and shapes the electronic band structure in
such a way as to produce further peculiar properties.

\section{Acknowledgments.}

We acknowledge discussions with J. Flouquet and J.P. Brison, and support from COST P-16 and from NES.
The Laboratorio de Bajas Temperaturas is associated to the ICMM of the CSIC. This work was supported
by the spanish MEC (Consolider Ingenio 2010 and grant FIS-2004-02897), and by the Comunidad de Madrid
through program "Science and technology at the millikelvin" (S-0505/ESP/0337).

\end{document}